# ABOUT THE POSSIBLE SIGNATURE OF PION BOSE CONDENSATION CONTRIBUTION IN THE HIGH ENERGY COSMIC RAY NUCLEAR INTERACTIONS

T.T.Barnaveli*(1), T.T.Barnaveli (jr) (1), A.P.Chubenko (2), N.A.Eristavi (1), N.M.Nesterova (2) and I.V.Khaldeeva (1)

(1) Andronikashvili Institute of Physics, Tamarashvili 6, Tbilisi, Georgia
(2) Lebedev Physical Institute, Leninsky av. 53, Moscow, Russia

* E-mail: tengiz.barnaveli@gmail.com

**ABSTRACT**

The events of anomalous energy deposit and high number of hadrons in hadron calorimeter, often not accompanied at all, or accompanied by very small EAS of cosmic rays, are observed by means of the complex installation aimed at research of cosmic rays. The data quoted and their analysis point to the real contribution and the essential role of pion Bose condensation in high energy nuclear interactions. The possibility of the existence of such process in the nuclear matter (in the neutron stars and in some kinds of the nuclei) was proposed by A.B.Migdal in 1971 [1] .

**THE EXPERIMENTAL DATA**

The present work is based on the analysis of archive experimental data obtained at Tian-Shan integrated installation in 1970-s. The energy of EAS hadron component was measured by means of the multilayer hadron ionization calorimeter of the area of 36 $m^2$ . The complete description of installation one can find e.g. in [4]. The data treatment was carried out according the essentially updated algorithms elaborated in the last several years (see e.g. [5,6]). In the present work our analysis is based not on the incident EAS size, as it was done e.g. in the quoted works, but on the energy deposit in the hadron calorimeter.

Note that the installation was triggered by the transducers registering the EAS ("EAS trigger") as well as by the calorimeter ("calorimeter trigger"). In both cases the indications of all transducers of the installation were registered without exception.

Basing the analysis on the energy deposit in calorimeter, we have found a rather interesting phenomenon. It turned out that some events of high energy deposit and of high number of hadrons entering the calorimeter are accompanied by very small EAS ($N_e < \sim 300 - 400$), and sometimes EAS were not fixed at all. I.e. the indications of EAS detectors were so small that EAS did not submit to the detailed treatment, in the sense of sufficiently reliable determination of the particle number $N_e$, age parameter $S$ and coordinates X and Y of event axis falling point. Evaluation of these parameters just by the order of value was possible. Only the spatial angle of the event axis could be determined with sufficient accuracy. The events fixed in the calorimeter were rather clear and enabled such analysis.

As an example of such events the visual images of the energy deposits EK in calorimeter are shown in the figures 1 – 3. For EK the arbitrary units proportional to the total ionization in calorimeter are used. The both projections of the calorimeter are shown. Each projection is 6 m in width.

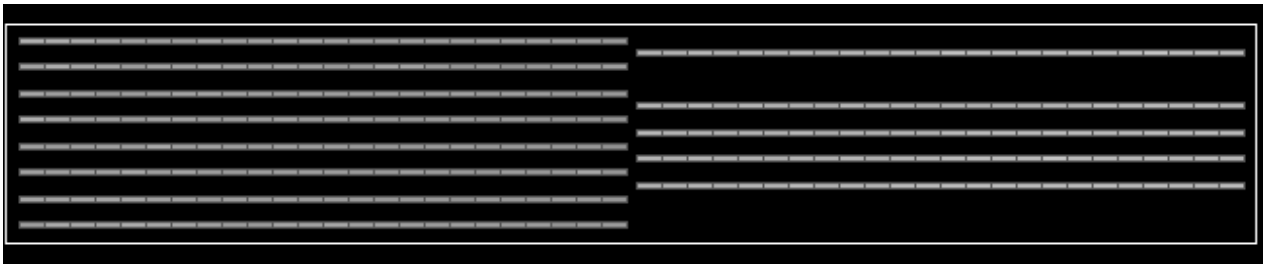

Fig. 1. Energy in calorimeter 20996.7, EAS size 336, zenith angle 43.53˚

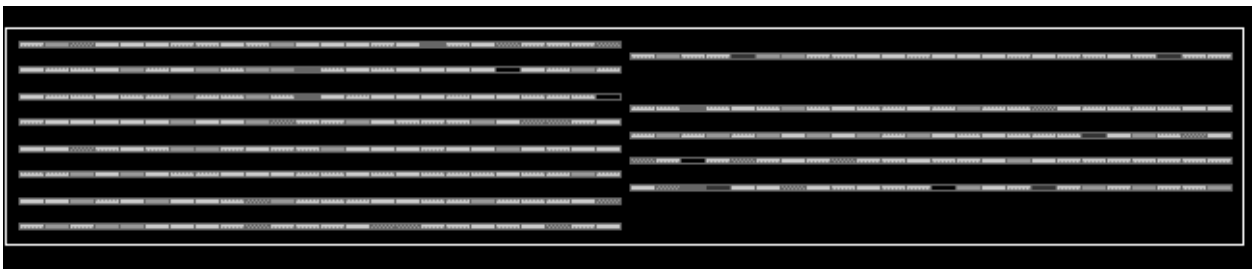

Fig. 2. Energy in calorimeter 981294.2, EAS size 344, zenith angle 37.39˚

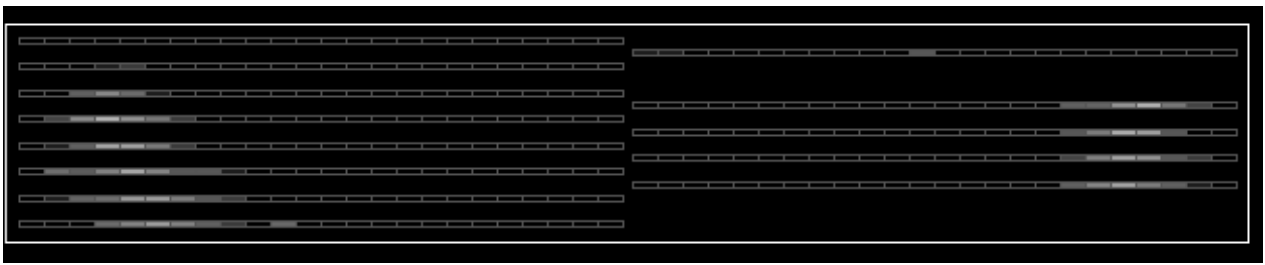

Fig. 3. Energy in calorimeter 5179.1, EAS size 356, zenith angle 19.43˚

We underline that these events are distributed over the whole volume of the handled experimental material accumulated in the period of about 3 years. The overwhelming majority of these events,∼ 95 %, are like the events in figures 1 and 2, i.e. the energy deposits in the calorimeter are caused not by the separate energetic hadrons. In reality the wide dense stream of hadrons cowers the calorimeter. The energy deposit takes place practically in all chambers of the calorimeter. Among these "anomalous" events just in a small part of them (< 5%) one can visually discern the separate cascades (Fig. 3). Seemingly, in this concrete case the axis of event fell at larger distance from the center of installation, i.e. of the calorimeter, and it was hit by the peripheral part of the bundle, nevertheless the "calorimetrical trigger" came into action.

Note as well, that in the "ordinary" events, when there exists the "full-bodied" accompanying EAS (fig. 4), the angle measured by means of calorimeter just slightly differs in average from that measured with the help of chronotron [5]. A long time experience of the work with these experimental data showed us that the angle measured with the help of calorimeter is even more reliable.

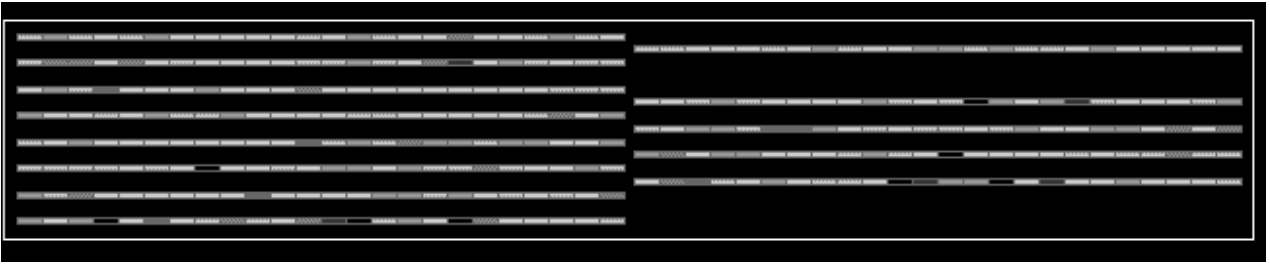

Fig. 4. Energy in calorimeter 804623.6,  EAS size 1.22 · $10^5$, zenith angle 51.26°

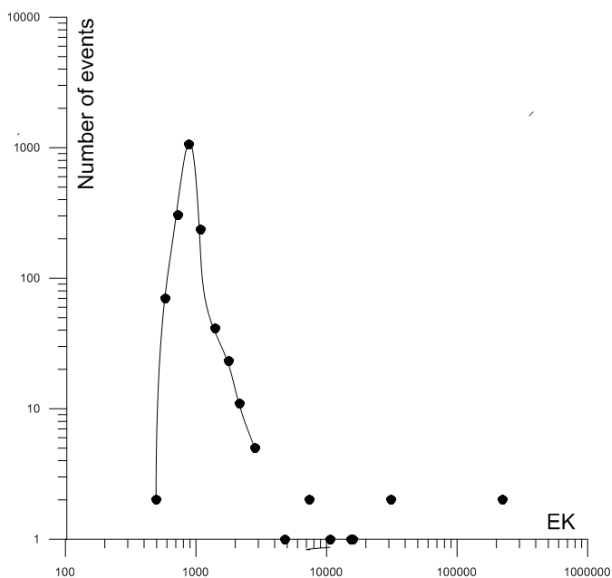

Fig. 5

After the discovery of the phenomenon under consideration the special selection was carried out of the events where the large energy deposit EK in the calorimeter was accompanied by so insignificant electron-photon cascade, that evaluation of its size (the number of particles) was possible just by the order of value. Such events build rather small part of the experimental data bank, < 0.1 %.  In the Fig. 5 the spectrum of such events is shown, i.e. the spectrum of energy deposits in calorimeter accompanied by very small electron – photon cascades, the characteristics of which could be evaluated just by the order of value (with exception of the spatial angle measured with calorimeter). The maximum of the spectrum is located at EK.~ 1000, the lower border of the spectrum EK~ 500.  We specially underline that at some values of energy exposures EK the events are absent – Fig. 5.

We emphasize the attention to the events without of the accompanying showers – exactly these events allowed to discover the phenomenon under consideration. Here the process takes place in its "pure state". However, as it will be shown below, this phenomenon may take place at

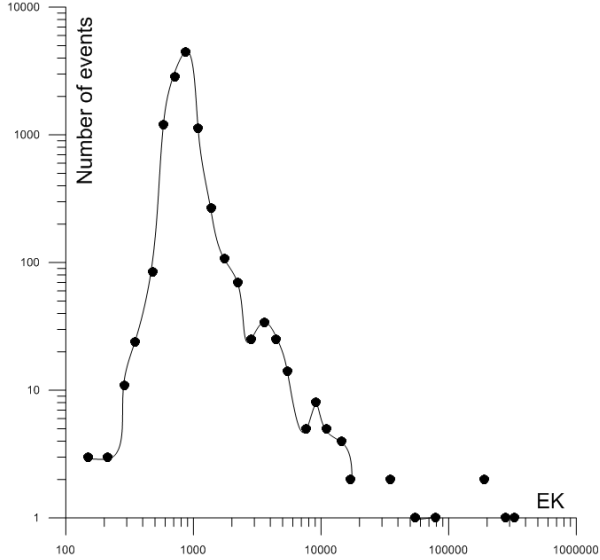

Fig. 6

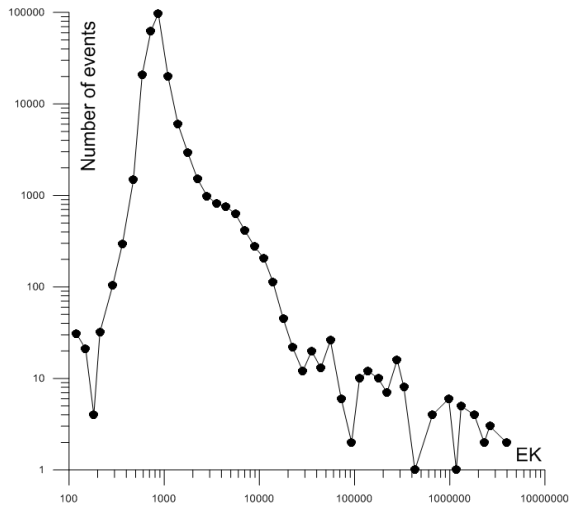

Fig. 7

any primary energy, though it becomes less noticeable against the background of the large powerful electron photon-nuclear cascades (EAS). As we will see, the special analysis will reveal the presence of this process at high energies as well. For the present, let us notice that such events could be created by very narrow bundles (cascades) penetrating through the calorimeter, but not touching the transducers located out of the central part of installation. I.e. actually this may be the electron – photon cascades of very small age parameter $S$. To check these considerations the spectrum of energy deposits for $N_e > 1000$, i.e. up to arbitrary large values, but of the age parameter $S < 0.1$, was built – Fig 6. Proceeding from the smallness of the parameter $S$ the majority of these events evidently were generated at small heights above the installation. Here the left border of the spectrum is lower, EK ~ 50. The complete spectrum of energy deposits in the calorimeter, not depending in particular on the age parameter $S$ is shown in the Fig. 7

Now let us consider the mutual correlations of three most important parameters, characterizing the phenomenon under consideration. These are 1) energy deposit EK in calorimeter, 2) the accompanying EAS size $N_e$, 3) the age parameter $S$ of the accompanying cascade. So let us consider 3 correlation dependencies : ($N_e$, $EK$), ($EK$, $S$) and ($N_e$, $S$).

These 3 correlation dependencies are shown in Figures 8, 9 and 10 correspondingly. The complete statistic is of about 100000 events, to each event corresponds 1 point in the graph.

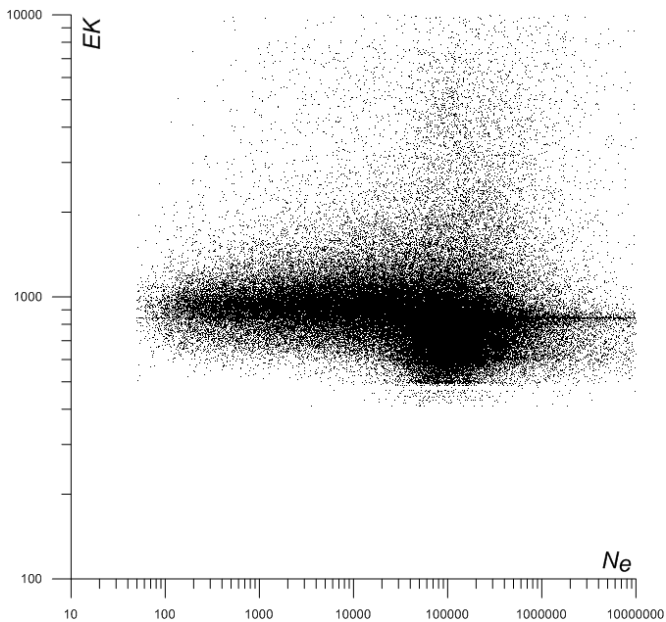

Fig. 8

In the figure 8 the correlation field ($N_e$, $EK$) is shown. Here one can clearly notice two regions of the event accumulation. They are adjoined to each other but clearly differ by the location of the weight center along the EK axis. These regions evidently are marking the events generated with preponderating of different mechanisms. The congestion in the right side of the graph – there are the events created mostly by the "traditional" mechanisms. The congestion in the left side – this is the region of very small $N_e$, those, the characteristics of which may be defined only by the order of value, as we have noticed above. Nevertheless the energy deposits here are very large. They are sufficiently higher in average then in the right side of the graph. Evidently in the majority of cases these events were generated at the small heights above the installation, from which the EAS had not enough time to develop. I.e. the age parameter $S$ of these events may be very small. Due to trifling $N_e$ the installation most probably was triggered by "calorimeter trigger". These events are as if "attracted" towards the center of the installation, i.e. to calorimeter. This part of events obviously was generated with participation of some "additional" mechanism.

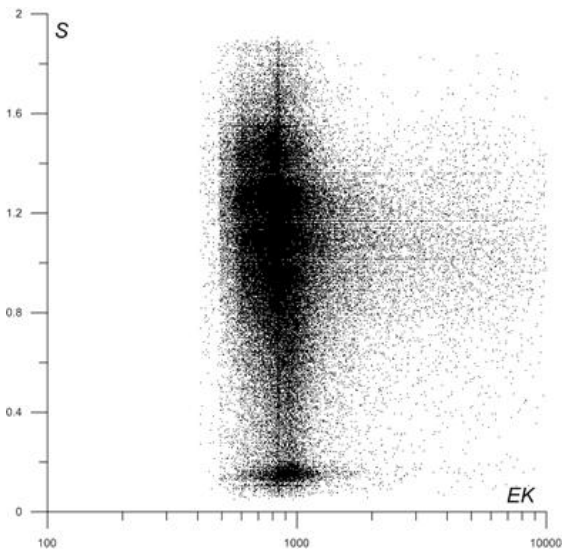

Fig. 9

In the figure 9 the correlation field ($EK$, $S$) is shown. Here the events generated by the "traditional" and "additional" mechanisms are already spatially separated from each other. This graph already points more definitely on existence of 2 different processes of event generation. Here the events of small parameter $S$ compose the separate conjunction without the smooth monotonous transition towards the region of high $S$ events accumulation. Here the conclusion on the existence of two independent processes of event generation suggests itself more clearly. One of these processes generates the events of small parameter $S$ distributed in the narrow diapason of values. The fraction of such events is small. In the main bulk of events the values of $S$ are distributed in the wide diapason of values. The both zones are distinctly separated from each other, being homogeneous inside themselves.

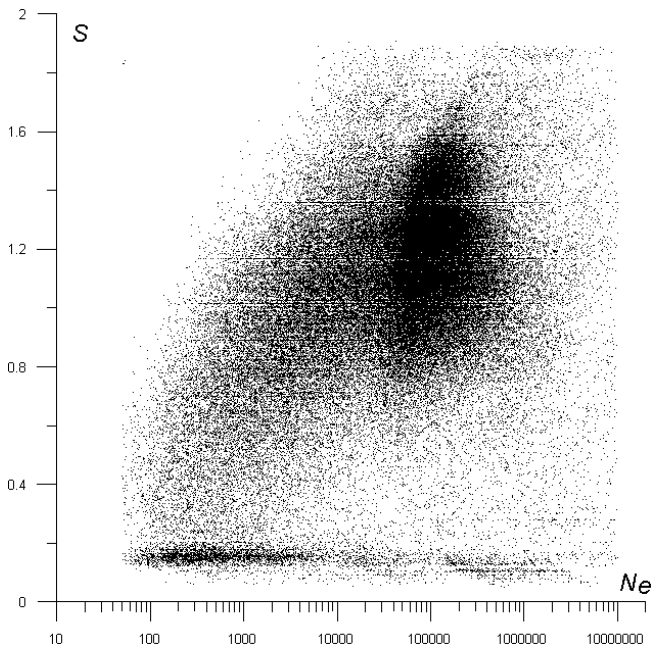

In this sense the next Figure 10, where the correlation field for ($N_e$,$S$) is shown, is very significant. Here again the 2 regions of event accumulation are distinguished distinctly separated from each other. One of them, most interesting for us, is the region of small $S$ located in the lower part of the graph. Part of these events had contributed to graph in Fig. 6. This structure may be traced along the whole range of $N_e$, i.e. of the primary energies.

Fig. 10

**DISCUSSION AND ANALISIS**

As to the nature of the "anomalous" events in the calorimeter it seems to us that most natural and prospective would be to describe them in the frames of pion condensation model, proposed by A.B.Migdal in 1971 [1]. Generally speaking, the questions of the particle spontaneous generation in different strong fields were discussed intensively in 1960 – 1979 – s. We'd like to mention especially the work of Ya.B.Zeldovich and V.S.Popov [3]. The list of references one can find e.g. in [1, 2].

The choice of this model is defined by the necessity to satisfy several indications established experimentally and presented above. **In particular :**

**At very small accompanying electron photon cascade:**
  a) The energy deposits in calorimeter are anomalously high.
  b) The events must contain a good developed hadron component, covering the calorimeter with broad dense stream.

**At large and very large accompanying electron photon cascade:**
  c) The noticeable number of cascades of very small age parameter $S$ must be observed.

**At any size of the accompanying electron photon cascade:**
  d) In the correlation diagrams the events of the small age parameter $S$ must be distinctly separated spatially from the other ones, without the smooth monotonous transition.

Among the alternative models the most attractive could be one based on the distribution of events according the heights of the primary particle interaction in the atmosphere. In such way one could explain the generation of "anomalous" events via primary particles which have managed to "slip" down practically the whole atmosphere without interaction. However such model does not correspond to the quoted sample of experimental data, in any case to items b), c) and d).

In the very schematic way the idea of the pion condensate emerge may be reduced to the following [1,2]. The gain of energy in the strong force field may supply the rest mass for the virtually born particle. In other words in the strong force field, in particular in the strong field of nuclear forces, the particles will be born and accumulated until they will not create the additional field, for example the coulomb field, which will prevent the process, and accumulation of particles will be stopped. So in the strong force field vacuum turned out to be reconstructed and gets unstable in regard to the spontaneous creation of the particles. If these particles are bosons then a high amount of them may be accumulated in principle. In case of presence of such process in nuclei the most advantageous obviously is the accumulation of $\pi^-$ mesons - so the pion Bose condensate emerges. The positive charge of the nucleus contributes to the accumulation and holding of the cluster of condensate $\pi^-$ mesons. According to e.g. [2] the accumulation both of $\pi^+$ and $\pi^0$ is possible as well.

In case of experimental data under consideration, the 2 versions of pion condensate presence in the formation of these events are possible. Which of them works in reality (may be the both of them) is not clear yet.

1. According to model [1] the pion Bose condensation may take place only in the special conditions of reconstructed vacuum existing in neutron stars, or in the hypothetical super heavy nuclei, or in hypothetical nuclei of anomalous ratio of the nucleus mass $M$ to its charge $Z$. So as a first version of "anomalous" events in calorimeter one can suppose that such nuclei really exist in the nature and are presented in the primary spectrum, let it be in a small amount. How not to quote here the final lines of A.B.Migdal's article [1]: "May be among the heavy nuclei of cosmic rays there are the super heavy nuclei with the ratio of charge to mass significantly differing from the ordinary nuclei. Who knows?"

2. According [1,2] in some "ordinary" heavy nuclei the conditions "sufficiently close" to such medium exist as well. Under the words "sufficiently close" one evidently has to mean that in such nuclei the virtual fulfillment of such conditions and the emerge of the condensate in the virtual form in them is not excluded. In any case there is nothing forbidding it.

Now let's imagine that such fluctuation did happen in some nucleus. As a result the nucleus will virtually contain the pion condensate in the field of its nucleons. After all, in the nucleus there were the conditions "close" to the required ones.

The interaction with the outer target particle ensures the transfer of energy and momentum into the pion condensate medium, as directly so via the core nucleons. The part of transferred energy ensures the transition of the condensate virtual particles into the real state. Such transition is possible if the transferred energy $E > n \cdot m_\pi$, where $n$ – is the number of the condensate particles transmitting into the real state. The rest of the transferred energy gets redistributed between the condensate medium and the nucleons of the core. Exactly this way as well the momentum and in particular the transmitted transverse momentum is redistributed inversely proportionally to $n$ and the number of nucleons of the core. This is very important for the further formation of the picture of what is happening.

All the subsequent takes place in the way as if the condensate would be real from the very beginning. I.e. the same way, as it would happen in the case of the first version, for the super heavy nuclei or for those of the anomalous ratio $M/Z$. As a result of interaction the symmetry of the system gets upset. The particles of condensate medium will be transited in the new energetic states, the different ones for all of them, generally speaking. Therefore this medium cannot more remain to exist as a condensate and decays on the separate particles of small transverse moments, in average inversely proportional to the number of condensate particles (see above). These particles continue to move in the direction of the primary nucleus as a more or less compact bunch, true expanding one due to mutual repulsion of $\pi^-$ mesons. Incidentally, these forces are compensated by the positive charge of core nucleons to a great extent. Due to the quite evident considerations these charges (of the condensate and of the nucleus) must be of the same order by value. Here one must not forget, that the parallel condensation of $\pi^+$ and $\pi^0$ mesons is possible as well.

Each particle of the bunch now carries the corresponding part of the primary nucleus energy. Entering the calorimeter, this bunch causes the large energy deposit. Let us underline once more that in these events the energy deposit in calorimeter is caused not by the single, or couple or three separate hadrons. In reality the calorimeter enters the wide dense bundle of hadrons. Evidently, the event occupies the area essentially larger, then the area of the calorimeter. To tell the exact number of the particles entering the calorimeter is difficult, because it is not a device showing the tracks of each particle. However taking into account that each layer of calorimeter consists of 24 ionization chambers and all of them are fixing the very high energy deposit, the number of hadrons may be recon by many tens if not by hundreds of particles. This will require the additional analysis but the order of numbers is exactly like that. Exactly such picture must be awaited in the frames of the developed model based on the pion condensate disintegration. With account of a small average transverse momentum in the bunch this process can form the described events even from the large enough heights.

The energy deposit in calorimeter is defined not only by the energy of separate particles of the bunch (it is of the same order for all of them), but also by the number of these particles (the probability to hit the calorimeter !), i.e. by the degree of condensate development. This last in its turn depends on the mass and structure of the "mother" nucleus. It is possible that the nuclei of a symmetric structure (of the equal number of P and N in nucleus) better stimulate the virtual creation of the condensate. The effectiveness of the process must increase with the growth of nucleus mass. As a result one may expect the emerge of dips and pikes in the EK spectrum. One can notice the hint pointing to such behavior in Figures 5 and 6. Due to smallness of the energy per hadron the development of the accompanying electron - photon cascade at the relatively small primary energies is suppressed.

One can carry out some analogy of the described process with the multi-peripheral model. In both cases the interaction with outer particle destroys the virtual picture. Introducing the energy into the system the interaction turns this system into the real one.

Note once more that pion condensation process (including the virtual one) may take place only in separate kinds of nuclei. Spectrum in Fig. 5 concerns to the events generated exactly with participation of pion condensation in its "pure state", practically not accompanied by the electron–photon cascade. Here the multiplicity of hadrons in calorimeter is high exactly due to the participation of this process. Of course the process of pion condensate generation may take place at any stage of primary nucleus passing through the atmosphere.

Let us return to the analysis of experimental data. Let us consider the correlation dependencies ($N_e$, $EK$), ($EK$, $S$) and ($N_e$, $S$). These 3 correlation dependencies were shown in figures 8, 9 and 10 correspondingly. In all these graphs one can notice all above mentioned peculiarities of the interaction with account of pion condensation contribution.

In the figure 8 one can notice two regions of the event accumulation. They are adjoined to each other but clearly differ by the location of the weight centrum along the EK axis. As it was said above, the congestion in the left side – this is the region of very small $N_e$. Nevertheless the energy deposits here are very large. They are sufficiently higher in average then in the right side of the graph. If these events were generated with participation of pion condensate then increased deposit of energies is easy to be explained. Really, as it was said above, the flux of hadrons created with participation of pion condensate increases due to small transverse momentum. Increase of the hadron flux naturally leads to increase of energy deposit.

In the figure 9 correlation field for ($EK$, $S$) is shown. Here again the 2 regions of accumulation are distinguished but already distinctly separated spatially from each other. One of them, most interesting for us, is the region of small $S$ located in the lower part of the graph. The small $S$ are distinguished because if the events are created with the participation of condensate then the hadrons will move as a compact bundle, moving close to the axis of event.

In the figure 10 the regions of event accumulation are spatially separated from each other even more clearly. Here the events of small $S$ are stretched along the whole range of $N_e$ up to the highest values. The point is that for the sufficiently high primary energies the hadron bundle, created from condensate decay, as well gives rise to the electron-photon component exactly in the vicinity the of EAS axis. As a result the spatial distribution of particles near the EAS axis gets very steep, what is characteristic for the cascades of the small age parameter $S$. So we can conclude that for the large values of $N_e$ the small $S$ are imitated due to the contribution of pion Bose condensation process. We have noticed above that the virtual process of pion condensation may take place at any height above the installation, i.e. at any stage of passage through the atmosphere. Exactly this region of the small $S$ events marking the contribution of the pion Bose condensation process can be traced along the whole lower part of correlation field up to the most large $N_e$. However, towards the large values of $N_e$ it gets more and more diffusive - because it gets more difficult for the process of condensation to distinguish against the background of large powerful "standard" electron-photon-nuclear cascades. Here the additional role belongs to the growth of the interaction cross section with the growth of the primary nucleus energy, i.e. with the growth of EAS size. The primary interaction takes place at the more and more high levels in the atmosphere so the electron-photon-nuclear cascade has enough time to reach the complete development. In the formation of cascade now the hadrons of condensate are participating as well, carrying increasing share of energy and besides they have enough time to spread wider around the axis of EAS.

## CONCLUSIONS

The quoted experimental data and their analysis speak for the real existence and contribution of the pion condensation process and its essential role in the high energy nuclear interactions. We think that this phenomenon deserves the further intensive investigation.

## ACKNOWLEDGEMENTS

The authors express their sincere gratitude to O.V.Kancheli for good-wishing critical remarks and valuable advises. We are grateful to the firm "Manta Systems" (Tbilisi) the computing capacities of which were used at the primary treatment of the experimental data.